\documentclass[aps,prl,superscriptaddress,twocolumn,floatfix,citeautoscript,longbibliography,hyperlinks]{revtex4-1}

\usepackage{graphicx}
\usepackage{dcolumn}
\usepackage[dvipsnames]{xcolor}
\usepackage[colorlinks,citecolor=blue,urlcolor=blue]{hyperref}
\usepackage{amssymb,amsmath,bm,bbm,mleftright,mathtools}
\DeclareMathAlphabet\mathbfcal{OMS}{cmsy}{b}{n}

\newcommand{\ket}[1]{\left|#1\right\rangle}

\makeatletter
\newsavebox{\@brx}
\newcommand{\llangle}[1][]{\savebox{\@brx}{\(\m@th{#1\langle}\)}%
  \mathopen{\copy\@brx\kern-0.5\wd\@brx\usebox{\@brx}}}
\newcommand{\rrangle}[1][]{\savebox{\@brx}{\(\m@th{#1\rangle}\)}%
  \mathclose{\copy\@brx\kern-0.5\wd\@brx\usebox{\@brx}}}
\makeatother

\begin{document}

\title{
Exponential shortcut to measurement-induced entanglement phase transitions
}

\author{Ali G. Moghaddam}\email{Email: agorbanz@iasbs.ac.ir}
\affiliation{Computational Physics Laboratory, Physics Unit, Faculty of Engineering and
Natural Sciences, Tampere University, FI-33014 Tampere, Finland}
\affiliation{Helsinki Institute of Physics, FI-00014 University of Helsinki, Finland}
\affiliation{Department of Physics, Institute for Advanced Studies in Basic Sciences (IASBS), Zanjan 45137-66731, Iran}
\author{ Kim P\"oyh\"onen}
\affiliation{Computational Physics Laboratory, Physics Unit, Faculty of Engineering and
Natural Sciences, Tampere University, FI-33014 Tampere, Finland}
\affiliation{Helsinki Institute of Physics, FI-00014 University of Helsinki, Finland}
\author{Teemu Ojanen} \email{Email: teemu.ojanen@tuni.fi}
\affiliation{Computational Physics Laboratory, Physics Unit, Faculty of Engineering and
Natural Sciences, Tampere University, FI-33014 Tampere, Finland}
\affiliation{Helsinki Institute of Physics, FI-00014 University of Helsinki, Finland}

\begin{abstract}
Recently discovered measurement-induced entanglement phase transitions in monitored quantum circuits provide a novel example of far-from-equilibrium quantum criticality. Here, we propose a highly efficient strategy for experimentally accessing these transitions through fluctuations. Instead of directly measuring entanglement entropy, which requires an exponential number of measurements in the subsystem size, our method provides a scalable approach to entanglement transitions in the presence of conserved quantities. In analogy to entanglement entropy and mutual information, we illustrate how bipartite and multipartite fluctuations can both be employed to analyze the measurement-induced criticality. Remarkably, the phase transition can be revealed by measuring fluctuations of only a handful of qubits.
\end{abstract}

\maketitle

\emph{Introduction}.---Entanglement is not only the most counterintuitive feature of quantum mechanics but also serves as the key resource of quantum information and the cornerstone of quantum technologies \cite{Horodecki2009, plenio2014}. Intriguingly, in isolated many-body systems that are in an excited state or driven under unitary dynamics, we typically find an abundance of entanglement \cite{Vedral2008review, Rigol2018, Barthel2021prl, Barthel2022}. In fact, starting from a state with no or low entanglement, unitary quantum dynamics governed by a generic many-body Hamiltonian inevitably generates entanglement. As a consequence, the system approaches a highly entangled state in which the entanglement entropy of a given subsystem scales with the subsystem volume, in sharp contrast to the area-law behavior typical of the ground states of many-body systems. This phenomenon is best illustrated in random quantum circuits where the successive application of local unitary gates gives rise to linear growth in entanglement \cite{Nahum2017}.

Quite recently, it has been found that the interplay between local measurements and unitary dynamics in monitored quantum circuits results in a critical behavior where, at a certain measurement rate $p =p_c$, a transition from volume- to area-law entanglement takes place \cite{Fisher2018,Fisher2019,Skinner2019,Smith2019,Vasseur2020,Huse2020,Huse2020PRL,Fisher2022review,potter2022review,DeLuca2019,Altman2020,Pixley2020,Zhu2020,Ashida2020,Chen2020,Buchler2020,Ruhman2021,lavasani2021measurement,Hsieh2021,Pal2021,Schomerus2021,Gullans2021,Diehl2021,Fazio2021,Khemani2021,Fisher2021,Vishwanath2021,Barkeshli2021,Hafezi2021,Lamacraft2022,Vasseur2022,Yao2022}.
Experimental endeavors have shown that observing this measurement-induced phase transition in quantum circuits can be challenging, as probing entanglement dynamics requires elaborate measurement setups \cite{Monrea2022experimental,Minnich2022experimental}.
A fundamental impediment is that characterizing entanglement entropy through quantum-state tomography is, as a task, exponentially hard in the subsystem size, and hence, in practice, limited to only few qubits in real experiments \cite{Cramer2010tomography,Eisert2010,Greiner2015measuring,Brydges2019probing}.

\begin{figure}[t!]
    \centering
    \includegraphics[width=.99\linewidth]{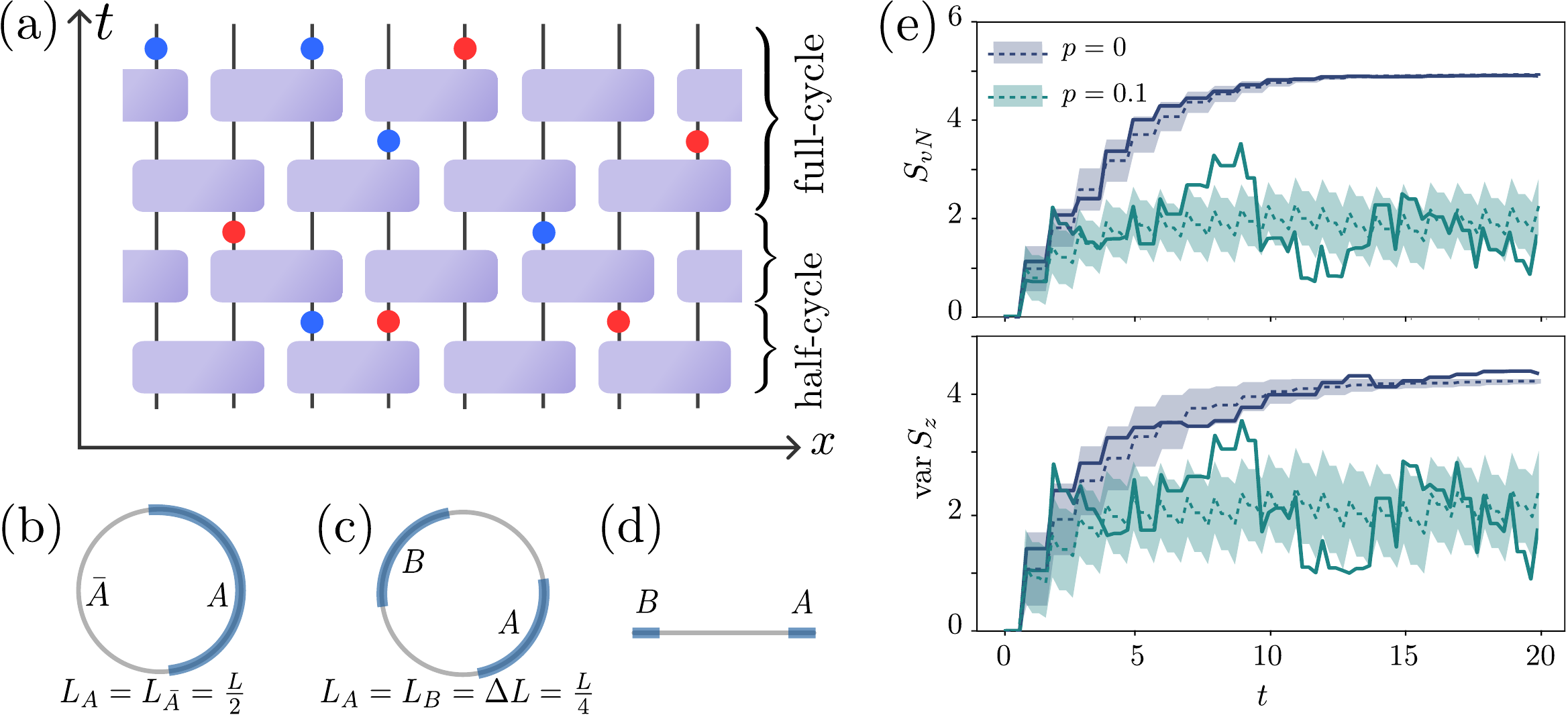}
    \caption{
    Schematic of the quantum circuit and its partitioning.
    In (a) each brick indicates a two-qubit unitary gate which are random under the charge-conservation constraint. Small circles represent single-qubit projective measurement of $s_z$ with red/blue colors corresponding to different measurement outcomes. (b) illustrates bi-partitioning of the system (grey circle) used to obtain entanglement entropy and spin variance of a subsystem. Similarly, (c) and (d) show the multi-partitioning of the circuit used to find mutual information and fluctuations between two subsystems indicated with thick blue lines. In (b) and (c) the circuit features periodic boundary conditions,
    while (d) corresponds to open boundary conditions removing the unitary gate connecting edges at the end of each half-cycle.
    (e) Time evolution of entanglement entropy and spin variance in a system with $L=16$ qubits and $L_s=8$ at different measurement rates. The solid lines show the results for a single trajectory while the dashed lines correspond to averaged result over 100 random trajectories, with the shaded regions representing the standard deviation around the average. 
    }
    \label{fig1}
\end{figure}

In this letter, we propose a scalable method to probe the entanglement dynamics and the measurement-induced phase transition.
This approach is motivated by the recently argued general connection between the scaling laws of entanglement entropies and variances of conserved extensive observables \cite{poyhonen2021}. Measuring variances, in contrast to the entanglement entropy, does not require a number of measurements exponential in the subsystem size. Considering a random quantum circuit with $U(1)$ charge conserving two-qubit gates, schematically shown in Fig.~\ref{fig1},
we demonstrate the equivalent behavior of the entanglement entropy and the variances in transient and steady states. Furthermore, we generalize previous ideas connecting bipartite entropies and fluctuations to multipartite systems. We demonstrate that, instead of the entanglement entropy and mutual information, entanglement phase transitions in systems with conserved quantities can equally well be analyzed through bipartite and multipartite fluctuations. 

\emph{Monitored quantum circuits}.---Our model consist of a prototypical quantum circuit containing $L$ qubits (or equivalently $L$ spin-$1/2$ states)
arranged in a one-dimensional chain. At each discrete time step, we perform a unitary evolution followed by set of random local measurements.
Unitary transformations are assumed to be decomposed in terms of
pairwise local two-qubit unitary operators each acting on neighboring qubits. The decomposition is staggered in time such that 
two-qubit gates at even and odd time steps act on even and odd links between neighbors, respectively, as depicted diagrammatically in Fig.~\ref{fig1}. 
Therefore, we can consider two successive time steps as one full cycle $t$. In the absence of measurement, it is represented by the unitary operator 
\begin{equation}
    {\bf U}(t) =\prod_{\text{odd } n}U_{n,n+1}(2t-1)\prod_{\text{even } n}U_{n,n+1}(2t)
\end{equation}
where $U_{n,n+1}(\tau)$ is a two-qubit unitary acting on neighboring qubits (at positions $n$ and $n+1$) at time step $\tau=2t\:\rm{or}\:2t-1$. To guarantee the $U(1)$ charge-conservation (or, equivalently, the conservation of the total spin $z$-component), these unitary gates are decomposed as 
\begin{equation}
    U_{n,n+1} = \begin{pmatrix}
    e^{i\varphi_{00}} & &\\
     &  e^{i\varphi_{11}} &\\
     && {\cal U}_{2\times 2}
    \end{pmatrix}
\end{equation}
in the basis $\{\ket{00}$, $\ket{11},  \ket{01}$ , $\ket{10}\}$. Here, ${\cal U}_{2\times 2}$ is a generic $2\times 2$ unitary matrix constituting of four independent phases \footnote{Please note that we use charge-conserving and spin-conserving interchangeably depending on whether we interpret the qubits as represented by spin-$1/2$ states (with basis $\ket{\uparrow}$ and $\ket{\downarrow}$) or charge states (with basis $\ket{0}$ and $\ket{1}$).}.
At each time step and for each pair of qubits, we select a two-qubit random unitary $U_{n,n+1}$ by picking up all six phases that parameterize it from a uniform random distribution.

\begin{figure}[t!]
    \centering
    \includegraphics[width=.99\linewidth]{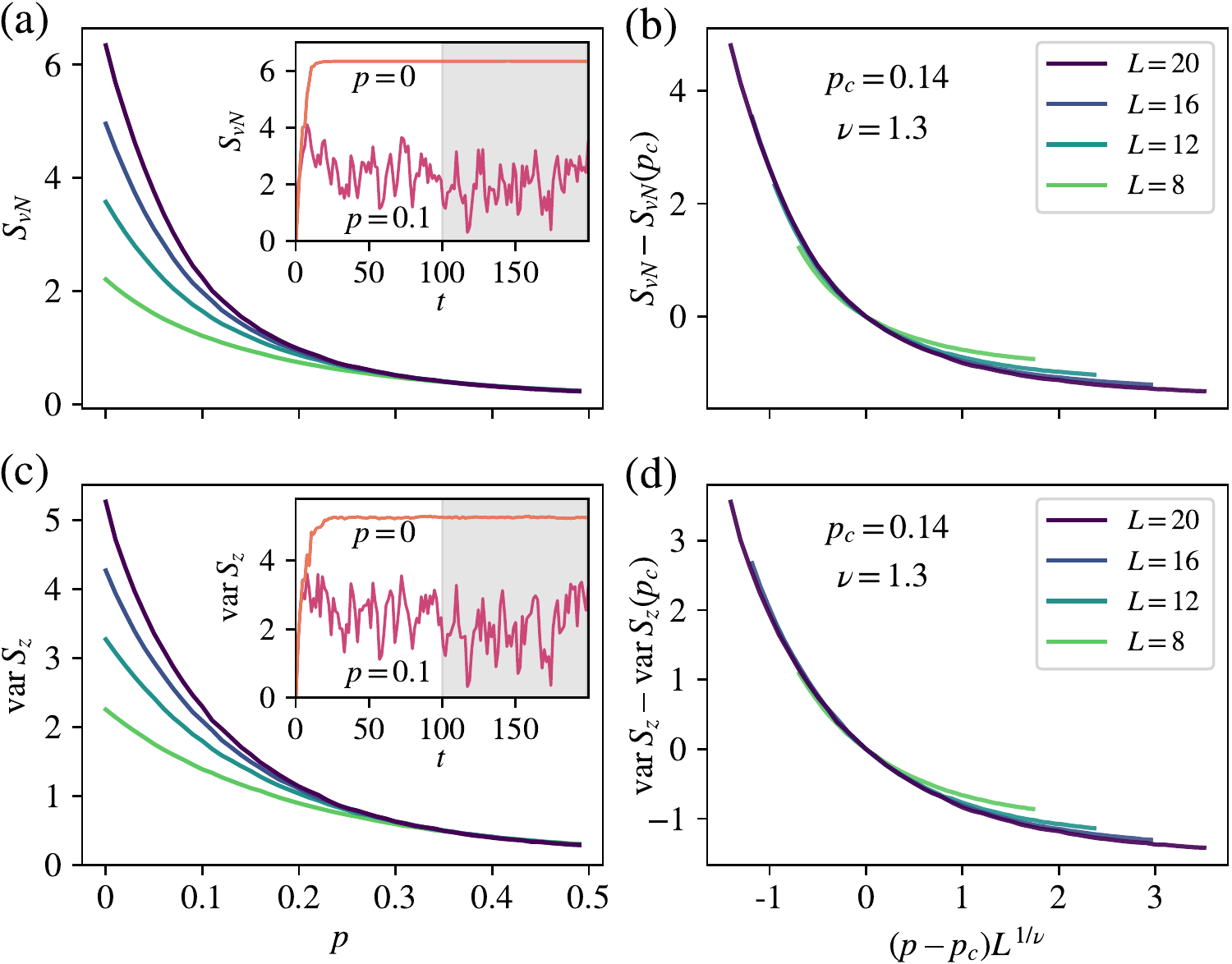}
    \caption{Equivalence of entanglement entropy and subsystem spin variance in monitored quantum circuits. (a) shows the averaged stationary value of entanglement entropy for half-sized subsystems ($L_s=L/2$) as functions of the measurement rate $p$ for different systems sizes $L$. 
    (b) shows the same result in a finite-size scaling form where we see data collapse for entanglement entropy of different system sizes.
    (c) shows the corresponding result for the spin variance 
    (d) also shows data collapse for the spin variance. 
    Insets of (a) and (c) show 
    the time evolution of entanglement entropy and spin variance for a single quantum trajectory at two different $p$'s.
    We observe indistinguishably similar behavior for entanglement entropy and spin variance as a function of measurement rate and also in their time evolution even at the level of a (unaveraged) single trajectory. The averages are taken over 100 full cycles
    from $t=100$ to $200$ (denoted in gray in the insets), and
    over ${\cal N} =240,300,450,600$ different realizations depending on the length $L=20,16,12,8$,respectively.
    }
    \label{fig2}
\end{figure}

As illustrated in Fig.~\ref{fig1}(a), the circuit dynamics is  accompanied by set of single-qubit measurement of randomly chosen qubits
at the end of each half-cycle. 
The measurement rate $p$ is defined such that 
at the end of each half cycle, each site has the probability $p$ of being measured. Hence, on average a fraction $p$ of all sites will be measured by the end of each half cycle.
Obviously, measurements make the full dynamics non-unitary and inherently probabilistic. Concentrating only on single-qubit projective measurements, the wavefunction randomly collapses
to an eigenstate of the corresponding single-qubit observable. 
The collapse is described by
\begin{equation}
    \ket\Psi \to \frac{1}{\lVert  P_{n,\alpha}\ket\Psi \rVert } P_{n,\alpha}\ket\Psi
\end{equation}
with projectors $P_{n,\alpha}$ acting on the qubit $n$ and satisfying normalization $\sum_{\alpha} P_{x,\alpha} = {\mathbb I} $. We focus on the case of measuring the $z$-component of the spin of the qubit whereby the projectors are given by $P_{n,\pm} = \big({\mathbb I} \pm s_{n,z}\big)/2$.
Each particular set of possible outcomes for the measurements
corresponds to a single quantum trajectory, which always remains a pure state. 

Using the instructions explained above, we can numerically evaluate the time evolution of full wavefunction of the quantum circuit for a given measurement rate. 
Then, from the wavefunction we can calculate entanglement entropy, variances, mutual information and mutual fluctuations in bipartite and multipartite settings, respectively, as schematically shown in Fig. \ref{fig1} (b-d).
The dynamics involves three independent sources of randomness: in the unitaries, in choosing the measured qubits, and in the sets of measurement outcomes. Accordingly, the expectation values of different quantities
can be obtained by taking their averages over all these forms of randomness.

\emph{Entanglement/fluctuation equivalence}.---To study the entanglement dynamics of the above model, we choose the initial state $\ket{\Psi} = \ket{\uparrow\downarrow\uparrow\downarrow \cdots \uparrow\downarrow}$, which, due to conservation, fixes the total spin $z$ to $S_z=0$ for all times. The entanglement phase transitions are defined in terms of 
the scaling behavior of the entanglement entropy 
as a function of subsystem length $L_s$.
This behavior does not depend on any particular definition of entropy, and here we focus on the von Neumann entropy
$S_{vN} = - {\rm Tr}\big(\rho_{s}\ln \rho_s\big)$, where $\rho_s$ is the subsystem reduced density matrix. As argued in our recent study \cite{poyhonen2021},
the fluctuations of conserved subsystem quantities are expected to exhibit the same dependence on the subsystem size as the entanglement entropy. To verify this principle for monitored quantum circuits, we analyze the fluctuations of the subsystem spin $\hat{S}_z = \sum_{n \in L_s} \hat{s}_{n,z}$ through its variance ${\rm var}\:S_z = \langle \hat{S}_z^2\rangle - \langle \hat{S}_z\rangle^2$. Indeed, as seen below, ${\rm var}\:S_z$ captures the behavior of $S_{vN}$ in great detail.

The unitary gates start to generate entanglement which initially, on average, grows linearly in time. After a time $t\sim L$, the system reaches a steady state, as illustrated in the inset of Fig.~\ref{fig2}(a). As seen in Fig.~\ref{fig2} (a), where we plot the trajectory-averaged entanglement entropy as a function of the measurement rate $p$ for different subsystem lengths, our findings for the $U(1)$ conserving model are qualitatively in line with the previously studied models without conservation. 
In particular, in the absence of measurements $(p = 0)$, we find that the steady-state wave function becomes maximally entangled, obeying an
entanglement volume-law scaling $S_{vN} \propto L_s$.
Here, however, due to the symmetry, the full system remains within a specific spin sector ($S^{\rm full}_z=0$) throughout its time evolution, and consequently, the maximal entanglement is also achieved within that sector.
The volume law persists at nonzero measurement probability up to a critical measurement rate $p=p_c$. Above the criticality $p>p_c$, the entanglement entropy follows an area law $S_{vN} \propto \ln \xi$, where $\xi$ is an effective \emph{system size independent} correlation length characterizing the spatial extent of entanglement. From previous works, we expect $\xi$ to be finite for $p>p_c$, and exhibit a power-law divergence at the criticality $\xi\propto (p-p_c)^{-\nu}$. As illustrated in Fig.~\ref{fig2} (b), this is supported by the fact that the entanglement data for different system sizes can be collapsed on a single curve with the standard scaling form $S_{vN} - S_{vN}(p_c) \propto f_{S}[(p-p_c)L^{1/\nu}]$ near the criticality \cite{Fisher2019}.

In Figs.~\ref{fig2} (c) and (d) we show the corresponding plots for ${\rm var}\:S_z $ instead of the entanglement entropy. Comparing with the results in Figs.~\ref{fig2} (a) and (b), one can verify the excellent match between the fluctuations and the entropy. 
Also, applying standard quantitative measures, we find both the Spearman's and Kendall's correlation coefficients are equal to 1, 
indicating a perfect monotonous relation between
${\rm var}\:S_z $ and $S_{vN}$ shown in Fig. \ref{fig2}(a) and \ref{fig2}(c).
Beyond the averaged quantities, even single quantum trajectories and their dynamics match well, as can seen by comparing the insets and inspecting Fig.~\ref{fig1}(e).
Importantly, fluctuations and the entropy both show a volume-law behavior for small $p$ and transition to the area-law phase at $p=p_c$. The equivalence between ${\rm var}\:S_z $ and $S_{vN}$ is yet more evident in the scaling collapses in Figs.~\ref{fig2} (b) and (d), which are identical apart from the numerical values on the vertical axis.

\begin{figure}[t!]
    \centering
    \includegraphics[width=.99\linewidth]{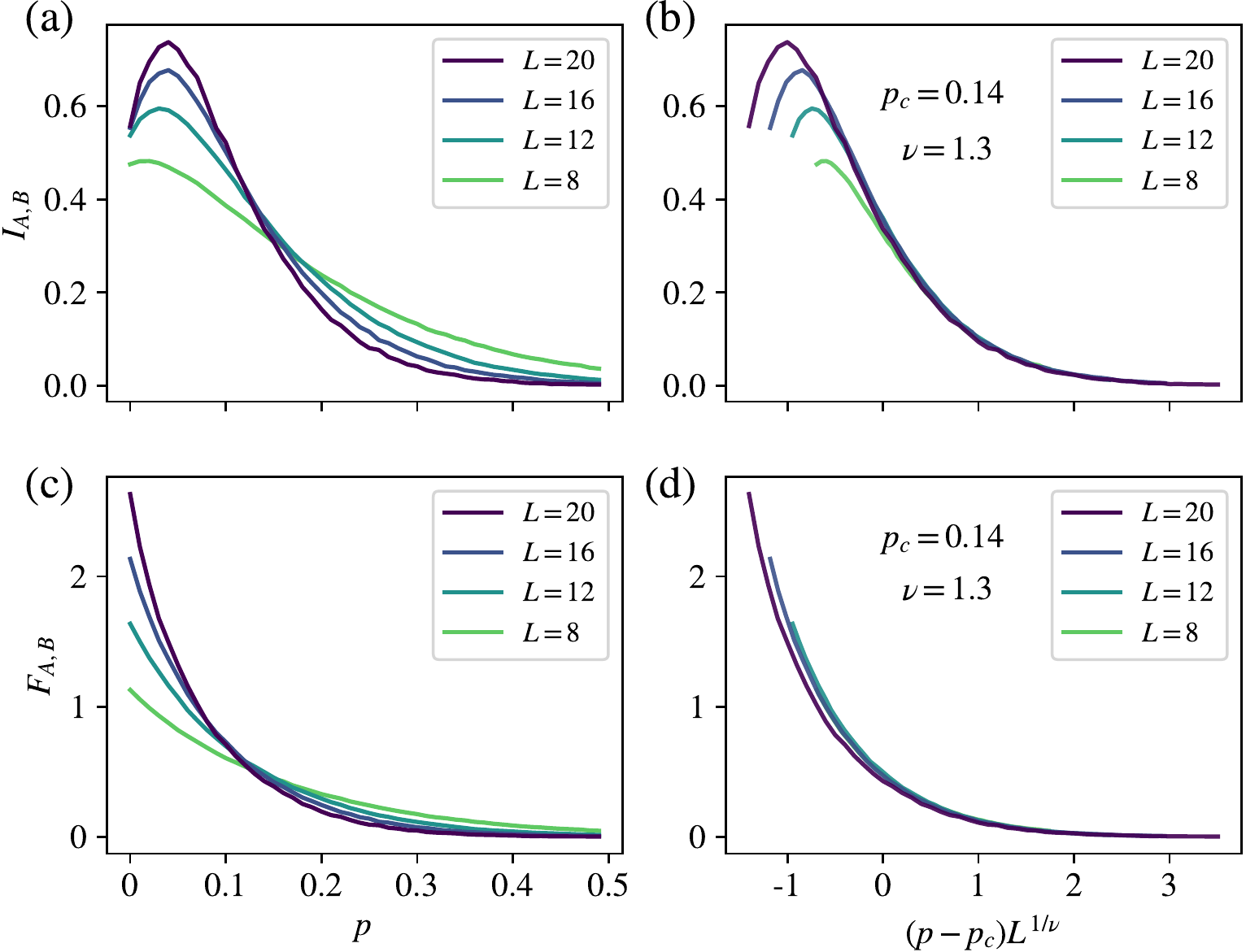}
    \caption{Mutual information $I_{A,B}$ and mutual fluctuations $F_{A,B}$ as functions of the measurement rate $p$.
    Panels (a) and (b) show mutual information between two quarter-sized disjoint subsystems ($L_{A}=L_{B}=\Delta L = L/4$ in non-scaled and scaled forms, respectively. Likewise, (c) and (d) show the behavior of mutual fluctuations. We have used the same values for the critical $p_c$ and correlation length exponent $\nu$ obtained from entanglement entropy (and spin variance) in Fig. \ref{fig2}, and we find again quite good data collapse especially around the critical point.
    }
    \label{fig3}
\end{figure}

\begin{figure}[t!]
    \centering
    \includegraphics[width=.99\linewidth]{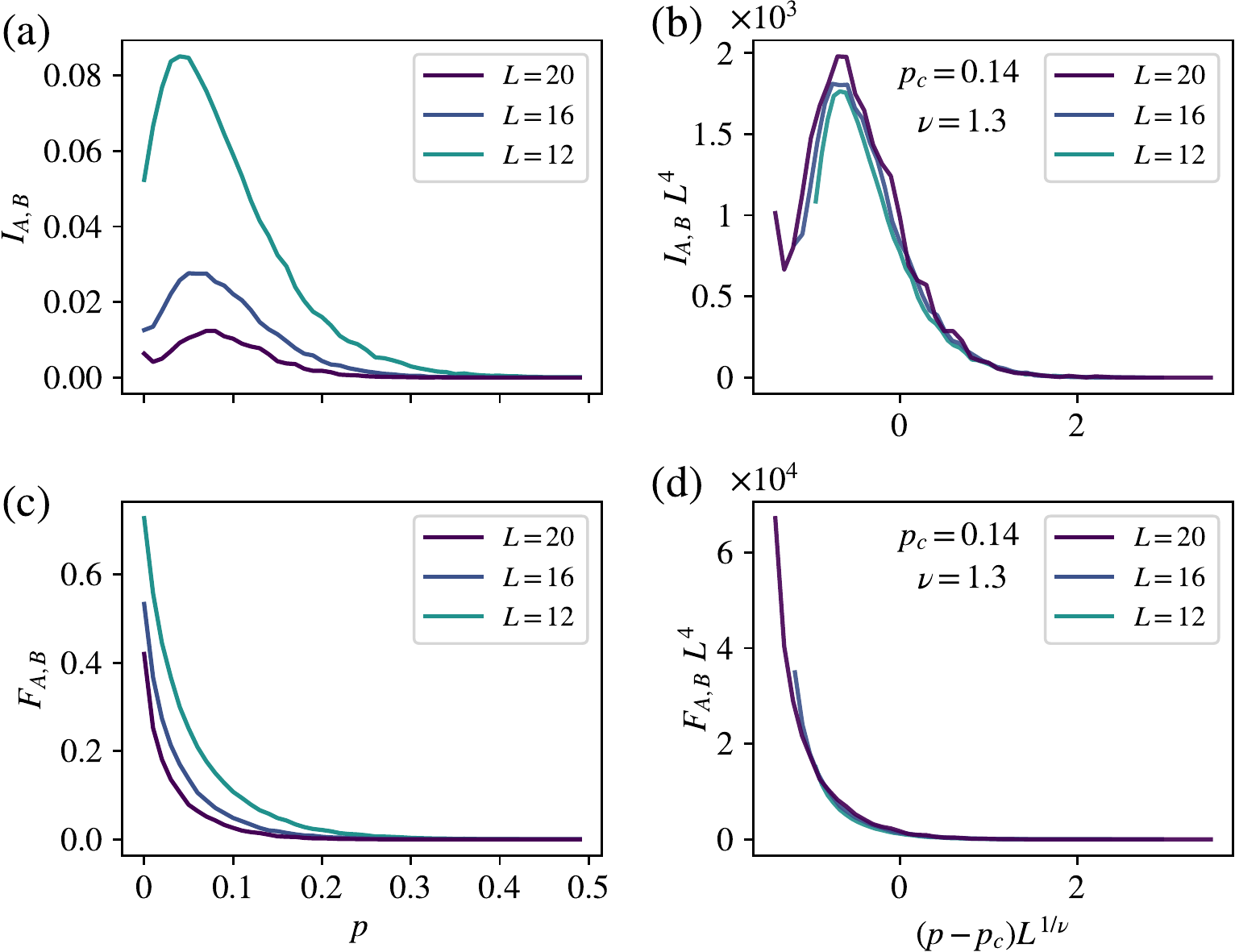}
    \caption{ Mutual information $I_{A,B}$ and mutual fluctuations $F_{A,B}$ as functions of measurement rate $p$ between two small-sized subsystems, each consisting of only two qubits in the partitioning geometry shown in Fig. \ref{fig1}(d).
    Panels (a) and (b) show the mutual information while (c) and (d) correspond to mutual fluctuations.   
    Apart from the differences in the detailed form of variations of $I_{A,B}$ and $F_{A,B}$ with $p$ in the behavior, both quantities show a $L^{-4}$ dependence on the length (or in other words the distance between the two subsystems) at the vicinity of critical point. This can be easily seen from the scaled results in panels (b) and (d) which also demonstrate the data collapse in the finite-size scaling analysis using the previously obtained values for $p_c$ and the exponent $\nu$.}
    \label{fig4}
\end{figure}

\emph{Multipartite entanglement and fluctuations}.--- Further information about entanglement in many-body systems can be obtained through multipartite entanglement measures.  A popular choice is the mutual information $I_{A,B}$, which quantifies quantum correlations between two disconnected subsystems $A$ and $B$. It is defined as $I_{A,B}=S_{A}+S_{B}-S_{A \cup B}$, where  $S_{\Omega}$ is typically, as here, taken as the von Neumann entropy.
As a measure of entanglement, the mutual information also behaves distinctively in the volume-law, critical, and area-law regimes \cite{Fisher2019,Vasseur2020}. 
 
Considering the intimate connection between bipartite entanglement and the fluctuations established in the last section, it is natural to study whether similar connection exists for multipartite entanglement measures and fluctuations. To this end, we compare the behavior of the mutual information $I_{A,B}$ and the corresponding mutual fluctuation $F_{A,B}={\rm var}\, S_{z,A}+{\rm var}\, S_{z,B}-{\rm var}\, S_{z,A\cup B}$ when the system is undergoing an entanglement phase transition. The results are presented in Figs. \ref{fig3} and \ref{fig4} for the partitioning illustrated in Figs. \ref{fig1}(c) and (d), respectively. When the size and separation of subsystems scale with the entire system length, both mutual properties exhibit a crossing near the transition between volume- and area-law regimes, as shown in Figs. \ref{fig3}(a) and (c). In the volume-law phase ($p<p_c$), the behavior of mutual properties grows with the length, while in the area-law phase, they decrease with an increase in $L$. At the transition point $p=p_c$, the length dependence disappears, giving rise to a crossing of the curves. Assuming a finite-size scaling relation $I_{A,B},F_{A,B}\propto f_{I,F}[(p-p_c)L^{1/\nu}]$ and using the same critical parameters found earlier, we obtain an excellent data collapse near criticality, as seen in Fig.~\ref{fig3} (b) and (d).

Finally, we consider the mutual information and fluctuations between two small subsystems, each consisting of two qubits located at the two ends of a system with open boundary conditions, as seen in Fig. \ref{fig1} (d). The results in Fig. \ref{fig4} show that the explicit dependence of $I_{A,B}$ and $F_{A,B}$ for small $p$ seems qualitatively different. Nevertheless, a closer inspection reveals that both quantities are seen to exhibit a $L^{-4}$ dependence in the vicinity of the transition, and an exponential suppression deep in the area-law phase. This is in agreement with previous studies, which found that the mutual information of small subsystems falls off as $L^{-4}$ at the criticality \cite{Fisher2019}. Thus, despite the superficial differences, as seen in Figs.~\ref{fig4} (b) and (d), the mutual properties for small subsystems can both be collapsed by the same finite-size scaling form $I_{A,B},F_{A,B}\propto L^{-4}f_{I,M}[(p-p_c)L^{1/\nu}]$ with the same critical data that was employed previously.
We note that the maxima in Figs.~\ref{fig3}(a) and \ref{fig4}(a) are a consequence of entanglement monogamy: the entanglement between two subsystems in a larger system can only be increased by reducing the entanglement with their complement \cite{Wootters2000,Winter2004}. This explains why at $p=0$, when the subsystems are maximally entangled with their complement, $I_{A, B}$ is suppressed. On the other hand, at $p\gg p_c$ the mutual information is suppressed because the system is weakly entangled. Thus, there exists a small but finite $p$, where $I_{A, B}$ obtains its maximum.

\emph{Discussion}.---In the pioneering works \cite{KlichLevitov2009,LeHur2010,LeHur2012},  the variance of conserved quantities was observed to qualitatively reflect the behavior of the entanglement entropy 
for area-law and critical ground states of free and interacting Hamiltonians. Building on these works, Ref.~\cite{poyhonen2021} provided a general state counting argument, according to which fluctuations of extensive conserved observables exhibit the same size-scaling laws as the entanglement entropy. In the present work we demonstrate this entanglement-fluctuation correspondence, for the first time, for volume-law and time-dependent  systems. Moreover, we generalize the relation between entanglement and fluctuations beyond the bipartite case.

Our work provides a clear prescription of how to access  entanglement phase transitions through fluctuations of conserved quantities. As an experimental probe of entanglement in quantum circuits, this could prove fruitful for several reasons. Entanglement entropy, which is generally used to define phase transitions in quantum circuits \cite{Fisher2019,Vasseur2020,Fisher2022review,potter2022review}, requires extensive knowledge of the the reduced density matrix of the subsystem whose size grows exponentially with subsystem size. Consequently, experiments measuring the entanglement entropy directly are, and will be, limited to very small systems \cite{Monrea2022experimental,Minnich2022experimental}.
On the other hand, probing fluctuations for a given many-body quantum state does not require full knowledge of the density matrix, and for a subsystem of $N$ qubits, ${\cal O}(N)$ number of $N$-qubit measurements will be sufficient similar to the sampling variance for a classical random variables with order $N$ possible outcomes. 
One key point is that, to probe the subsystem spin variance, we only need to measure these qubits in their natural basis (0 and 1). In contrast, probing entanglement directly through reduced density matrix tomography necessitates $2^N+1$ additional tomography circuits each for evaluating the outcome for one set of mutually commuting Pauli strings of the form $\sigma_{i_1}\otimes\cdots\otimes\sigma_{i_N}$ \cite{footnote1}.
Therefore, probing variance as an indirect measure of entanglement 
constitutes an exponential shortcut compared to the direct quantum tomography approach. For a more quantitative comparison, in a recent experiment \cite{Minnich2022experimental}, subsystems with sizes up to 5 have been considered, needing $2^5+1=33$ tomography circuits. However, for even slightly larger  subsystems of sizes $10$ and $20$ any single-shot measurement of the density matrix will require $10^3$, and $10^6$ different additional tomography circuits, respectively. Measuring fluctuations does not require these extra circuits and is thus much more feasible in existing setups with $\sim50$ qubits \cite{google2019quantumsupremacy}.
The only requirement is to have a sufficiently accurate control over the unitaries to implement the conservation,  which could already be achieved approximately \cite{Monrea2022experimental}, and is continuously improving \cite{Heinsoo2022PRXQuantum}.

\emph{Conclusions}.---We have proposed that the entanglement entropy in monitored unitary circuits can be accessed through the fluctuations of a conserved quantity, constituting an exponential reduction in the required measurements. We showed how this strategy can be employed to study non-steady-state entanglement dynamics as well as measurement-induced entanglement phase transitions through bipartite and multipartite fluctuations. In the simplest case, to reveal the existence of the transition, it is sufficient to measure fluctuations of only a handful of qubits.  Our results open new prospects to study the complex entanglement dynamics in existing and near-future quantum computers.

\bibliography{circuit_ms.bib}
\end{document}